\documentclass[12pt]{article}
\usepackage[super,compress]{cite}
\usepackage{amssymb}
\begin{document}

\begin{center}
{\bf Multiplicity distribution of gluons in pQCD}

G.H. Arakelyan$^1$,
Yu.M. Shabelski$^2$ and A.G. Shuvaev$^2$ \\

\vspace{.5cm}

$^1$A.Alikhanyan National Scientific Laboratory
(Yerevan Physics Institute)\\
Yerevan, 0036, Armenia\\
E-mail: argev@mail.yerphi.am\\

\vspace{.2cm}

$^1$Petersburg Nuclear Physics Institute, Kurchatov National
Research Center\\
Gatchina, St. Petersburg 188300, Russia\\
\vskip 0.9 truecm
E-mail: shabelsk@thd.pnpi.spb.ru\\
E-mail: shuvaev@thd.pnpi.spb.ru

\vspace{1.2cm}

\end{center}

\begin{abstract}

\noindent
The multiplicity distribution of the gluons
produced at the high energy is evaluated
in BFKL approach. The distribution has Poisson
form that can explain experimentally observed
KNO scaling.
\end{abstract}

\noindent
{\bf 1.}The multiple secondary particles production
at high energy hadron-hadron collisions
is known to exhibit KNO scaling~\cite{KNO}
confirmed in many experimental
papers~\cite{alner,ansorge}.
The scaling means that the ratio
$P_n(s)=\sigma_n(s)/\sigma_{in}(s)$
of the cross section to produce $n$ secondaries
in the collision event to the total inelastic
cross section depends on the energy $\sqrt s$
only through the averaged multiplicity $\overline n$,
$P_n(s)\,=\, 1/\overline n\,\Psi(n/\overline n)$.

According to Reggeon calculus the high energy
hadron scattering is mediated by the Pomerons
exchange, the Pomeron being described by
the cylinder type diagrams~\cite{CMV, Kaidalov:1982xe}.
The inelastic amplitudes come about from the $s$-channel
cutting of one or several Pomerons~\cite{Abramovsky:1973fm}.
Each cut can be associated with the creation of two
quark-gluon strings~\cite{Kaidalov:1983vn}
whose subsequent hadronization gives rise
to the multiple hadron production.

The observed KNO scaling has been
explained ~\cite{Kaidalov:1983vn, Kaidalov:1983ew}
by the various number of cut Pomerons (quark-gluon
strings) that may contribute to the interaction
amplitude.
Assuming Poisson distribution for the secondaries
produced from a single Pomeron the net sum over
the cuts yields the function $\Psi(n/\overline n)$.
In general it is not of the Poisson type,
its form is much more complicated.
It is nevertheless $s$ independent in the rather wide
energy interval although the scaling should be violated
at the very high energies in agreement with
the experimental data~\cite{Kaidalov:1983ew,
Hegyi:2000sp}.

There has been much theoretical activity in studying
KNO scaling in multiplicity distribution of QCD parton cascades.
The cascade starts off a primary virtual gluon
whose virtuality $Q^2$ is of the order of the typical
hard scale for the process, say, $e^+e^-$ annihilation.
The primary gluon emits secondary gluon jets each of which,
in turn, emits another jets and so on. The cascade development
is described by the evolution equation that collects
the leading or subleading powers of $\log Q^2$
(see e.g. \cite{Dokshitzer:1993dc}, or
\cite{Dokshitzer:1991wu} for review).

The task we are going to deal with here is alternative
in that we treat the secondaries arising from the large $s$
scattering of almost on shell particles rather than from
the decay of an virtual object. That is why the context
of Regge theory, collecting the powers of $\log s$ or,
equivalently, $\log 1/x$ for the deep inelastic scattering,
seems to be quite natural.
An important point to be questioned at first is
what kind of distribution could be expected from
a single Pomeron, or in another language, from a single
quark-gluon string. Is it evident that it is Poisson like?
Some insight can be gained by calculating
the distribution of perturbative gluons originating
from BFKL Pomeron.
It is the issue that is addressed below.

\noindent
{\bf 2.}The BFKL Pomeron arises as a compound state
of two reggeized gluons in the sum of ladder
type diagrams~\cite{FKL, KLF1, KLF2, BL}.
The treatment essentially relies
on Multi-Regge form of many-particle amplitudes.
It provides an expression for
the amplitude to produce $n$ gluons or, more generally
$n$ jets, in the inelastic scattering,
$A+B \to A^\prime +B^\prime + n$,
at the large invariant energy $s=(p_A+p_B)^2$
and the kinematics, when the rapidity
intervals between the jets are much larger those
between the jets' constituents - multi-Regge kinematics
(see \cite{FF} and references therein).
This amplitude is actually nothing else as
the cut through the Reggeon.
Using $s$-channel unitarity it gives
the imaginary part of
the elastic scattering amplitude,
which value at $t=0$ is translated into
the total cross section,
\begin{equation}
\label{sAB}
\sigma_{AB}(s)\,=\,\frac 14 \frac 1{(2\pi)^2}
\int\frac{d^{\,2}q}{q^2}\int\frac{d^{\,2}l}{l^2}\,
\Phi_A(q)\langle\,q|\,G(Y)\,|l\,\rangle\,
\Phi_B(l).
\end{equation}
Here $\Phi_{A,B}$ are the color-singlet impact-factors,
$Y=\ln s/s_0$ ($s_0$ is an appropriate energy scale
usually assumed in the Regge theory to be of the order
of~1~GeV$^2$),
$G(Y)$ is the Green function of two interacting
reggeized gluons. It is given by the series
\cite{Fadin:2006bj, IFL}
\begin{equation}
\label{ser}
G(Y)\,=\,e^{\Omega Y}\,+\,\sum\limits_{n=1}^\infty
\int\limits_0^Y\, dy_1\,e^{\Omega(Y-y_1)}K_r
\int\limits_0^{y_1}\,dy_2\,e^{\Omega(y_1-y_2)}K_r\,
\cdots\,
\int\limits_0^{y_{n-1}}\,dy_n\,e^{\Omega(y_{n-1}-y_n)}K_r\,
e^{\Omega {y_n}},
\end{equation}
with the operators $\Omega$ and $K_r$ acting in the transverse
momentum space. The operator $K_r$ describes the emission
of real (on-shell) gluons,
\begin{equation}
\label{Kr}
\langle q|\,K_r\,|l\rangle \,=\,
\frac{g^2N_c}{(2\pi)^3}\,\frac{2}{(q-l)^2+m_g^2},
\end{equation}
the virtual corrections are collected in the gluons'
trajectories, $\Omega=\omega_1+\omega_2$.
They are equal for the total transferred momentum $q=0$,
$\omega_1=\omega_2=\omega$,
\begin{equation}
\label{om}
\langle q|\,\omega\,|l\rangle \,=\,
-\frac 12\frac{g^2N_c}{(2\pi)^3}\,
\int d^{\,2}k\,\frac{q^2}{k^2(q-k)^2+m_g^2}\,
\delta^{(2)}(q-l),
\end{equation}
the gluon mass $m_g^2$ being the infrared cutoff.

The formulae (\ref{Kr}), (\ref{om}) are relevant in
the Leading Logarithm Approximation (LLA) collecting
the leading powers of $\alpha_S\ln s/s_0$, where
$\alpha_S=g^2/4\pi$, $g$ is QCD coupling constant.
The dominant contribution to cross sections of hard
processes comes in LLA from multi-Regge kinematics,
when the emitted particles are single gluons. The situation
allowing for the emission of the jets made up of two
or more particles corresponds to the next
to the leading order corrections~\cite{Fadin:2006bj,
Fadin:2015zea}.

There is an equivalent form of the operators (\ref{Kr}), (\ref{om}),
valid for the small $m_g$~\cite{IFL},
\begin{equation}
\label{log}
K_r\,=\,\alpha\,\bigl[-\ln m_g^2x^2 + 2\ln 2 + 2\psi(1)\bigr],
~~~~\omega\,=\,-\frac 12\alpha\,\ln\frac{q^2}{m_g^2},
\end{equation}
$\alpha\equiv N_c\frac{g^2}{4\pi^2}$,
$\psi(z)\,=\,\frac{d}{dz}\ln\Gamma(z)$.
The variables $x_k$ are the transverse space coordinates
conjugated to the momenta $q_k$ (the indices $k=1,2$
label the vectors' component in the transverse space),
$q^2=q_1^2+q_2^2$, $x^2=x_1^2+x_2^2$.
These variables are treated in the formula (\ref{log})
as operators $\hat q_k$,
$\hat x_k=i\frac{\partial}{\partial q_k}$ acting
in the transverse space, $|l\rangle$ being the operator $\hat q_k$
eigenstates, $\hat q_k|l\rangle=l_k|l\rangle$,
$\langle q|l\rangle = \delta^{(2)}(q-l)$.
The operator $\omega =\omega(\hat q_1,\hat q_2)$
is diagonal in this basis, its matrix elements
are given in (\ref{om}), the expression (\ref{Kr}) presents
the matrix elements of the non diagonal operator
$K_r=K_r(\hat x_1,\hat x_2)$. Hereafter we omits hats above
the operators' symbols.

The series (\ref{ser}) is summed up as
\begin{equation}
\label{G(Y)}
G(Y)\,=\,e^{KY},~~~K\,=\,K_r+\omega_1+\omega_2,
\end{equation}
which is the operator form of the BFKL equation~\cite{IFL}.
The mass $m_g$ cancels out in the operator $K$
for the color singlet Pomeron channel (as is evident
from (\ref{log})).

The $N$-th order term in the expansion (\ref{ser})
describes the emission of $N$ real gluons.
To pick up its relative weight in $G(Y)$
we multiply it by an auxiliary variable $u^N$,
or in other words replace $K_r \to u\,K_r$.
Then the probability to emit $n$ real gluons reads
\begin{equation}
\label{Pn}
P_N\,=\,\frac 1{P(1)}\,\frac 1{N!}\,
\frac{\partial^{\,N}}{\partial u^N}|_{u=0}P(u),
\end{equation}
with the generating function $P(u)$
obtained by modifying the formula (\ref{sAB}),
\begin{equation}
\label{P(u)}
P(u)\,=\,\int\!\frac{d^{\,2}q}{q^2}\int\!\frac{d^{\,2}l}{l^{\,2}}\,
\Phi_A(q)\langle\,q|\,e^{Y(uK_r+2\omega)}\,|l\,\rangle\,
\Phi_B(l).
\end{equation}

To work out this expression we firstly rearrange the operator
appearing in $P(u)$ as
\begin{equation}
\label{K(u)}
K(u)\,=\,u K_r+2\,\omega\,=\,\alpha\,\bigl[u\,(-\ln x^2
-\ln q^2)\,-\,\overline u\,\ln q^2\,
+\,\overline u\,\ln m_g^2\,+\,2u\,
\bigl(\ln 2+\psi(1)\bigr)\bigr],
\end{equation}
$\overline u\equiv 1-u$,
and pass to the complex variables,
$$
q=q_1+iq_2,~~q^*=q_1-iq_2,
~~~~ q^2=q\,q^*,~~~
x^2\,=\,-4\frac{\partial}{\partial q}
\frac{\partial}{\partial q^*},
$$
the hermitian conjugate being
$\left[\frac{\partial}{\partial q}\right]^+
=-\frac{\partial}{\partial q^*}$.
Using the identity~\cite{deVega:2003ji, IFL}
$$
\ln q\,+\,\ln \frac{\partial}{\partial q}\,=\,
\frac 12\bigl[\psi\bigl(1+q\frac{\partial}{\partial q}\bigr)
+\psi\bigl(-q\frac{\partial}{\partial q}\bigr)\bigr]
$$
and its hermitian conjugate we arrive at the form
that falls into the holomorphic and antiholomorphic
pieces,
$$
K(u)\,=\,\alpha\,\frac 12u\,
\bigl[\psi(1+D)+\psi(-D)+\psi(1+D^*)+\psi(-D^*)\bigr]
\,+\,\alpha\,L(u),
$$
$$
D\,\equiv\,q\,\frac{\partial}{\partial q},~~~
D^*\,\equiv\,q^*\,\frac{\partial}{\partial q^*},
$$
$$
L(u)\,=\,-\overline u\,\bigl(\ln q + \ln q^* - \ln m_g^2\bigr)
\,+\,2\,u\psi(1).
$$
Given the commutator $\bigl[\ln q\,,\,D\bigr]\,=\,-1$,
it is easy to check another operator identity,
$$
e^{\frac{u}{\overline u}Q}\bigl[-\overline u\,\ln q\,\bigr]
e^{-\frac{u}{\overline u}Q}\,=\,
-\overline u\,\ln q\,-\,\frac 12 u\,
\bigl[\psi(1+D)+\psi(-D)\bigr]
$$
with the operator
$$
Q\,=\,\frac 12\,\ln\frac{\Gamma(1+D)}{\Gamma(-D)},~~~~~~
Q^*\,=\,\frac 12\,\ln\frac{\Gamma(1+D^*)}{\Gamma(-D^*)}.
$$
Putting it together with the same identity for
the antiholomorphic $\ln q^*$
results into the relation
$$
e^{\frac{u}{\overline u}(Q+Q^*)}\alpha L(u)
e^{-\frac{u}{\overline u}(Q+Q^*)}\,=\,K(u),
$$
that in turn amounts to
\begin{equation}
\label{QLQ}
e^{YK(u)}\,=\,
e^{\frac{u}{\overline u}(Q+Q^*)}e^{\alpha YL(u)}
e^{-\frac{u}{\overline u}(Q+Q^*)}.
\end{equation}

Now one has to plug the identity (\ref{QLQ})
into the matrix element occurring in the function
$P(u)$ (\ref{P(u)}). We do it in the following way
$$
\langle q|\,
e^{\frac{u}{\overline u}(Q+Q^*)}e^{\alpha YL(u)}
e^{-\frac{u}{\overline u}(Q+Q^*)}|\,l\,\rangle
$$
$$
\,=\,
\langle q|\,\underbrace{e^{\frac 12\alpha YL(u)}
\,e^{-\frac 12\alpha YL(u)}}_1
e^{\frac{u}{\overline u}(Q+Q^*)}
e^{\frac 12\alpha YL(u)}
e^{\frac 12\alpha YL(u)}
e^{-\frac{u}{\overline u}(Q+Q^*)}
\underbrace{e^{-\frac 12\alpha YL(u)}
e^{\frac 12\alpha YL(u)}}_1|\,l\,\rangle
$$
By elaborating on the identities
$$
e^{\pm\frac 12\alpha YL(u)} D
e^{\mp\frac 12\alpha YL(u)} \,=\,
D \pm \frac 12\alpha Y \overline u
$$
along with the same for $D^*$ and
$$
\langle\,q\,|e^{\frac 12\alpha YL(u)}\,=\,
\langle\,q\,|e^{-\frac 12 \alpha Y \overline u
\ln\frac{q^2}{m_g^2}+
\alpha Y u \psi(1)},~~~~
e^{\frac 12\alpha YL(u)}|\,l\,\rangle\,=\,
e^{-\frac 12 \alpha Y \overline u
\ln\frac{l^2}{m_g^2}+
\alpha Y u \psi(1)}|\,l\,\rangle
$$
the function $P(u)$ is brought to the form
\begin{eqnarray}
\label{P(u)f}
P(u)\!&=&\!e^{\,2\alpha Y  u\psi(1)}\!
\int\!\frac{d^{\,2}q}{q^2}\!\!
\int\!\frac{d^{\,2}l}{l^2}\,
\Phi_A(q)\left(\frac{q^2}{m_g^2}\right)
^{\!\!-\frac 12\alpha Y \overline u}
\!\!\Phi_B(l)\left(\frac{l^{\,2}}{m_g^2}\right)
^{\!\!-\frac 12\alpha Y \overline u} \\
&&\times \langle\,q\,|H(u,D,D^*)|\,l\,\rangle, \nonumber \\
\label{HuDD}
H(u,D,D^*)\!&=&\!\exp\left\{
\frac 12 \frac{u}{\overline u}\left[
\ln\frac{\Gamma\bigl(1-\frac 12L\,\overline u+D\bigr)}
  {\Gamma\bigl(\frac 12L\,\overline u -D\bigr)}
+\ln\frac{\Gamma\bigl(1-\frac 12L\,\overline u+D^*\bigr)}
  {\Gamma\bigl(\frac 12L\,\overline u -D^*\bigr)}
  \right.\right. \\
&&-\left.\left.\ln\frac{\Gamma\bigl(1+\frac 12L\,\overline u+D\bigr)}
  {\Gamma\bigl(-\frac 12L\,\overline u -D\bigr)}
-\ln\frac{\Gamma\bigl(1+\frac 12L\,\overline u+D^*\bigr)}
  {\Gamma\bigl(-\frac 12L\,\overline u -D^*\bigr)}
\right]\right\},~~~L\equiv\alpha Y. \nonumber
\end{eqnarray}
This form has a merit of the infrared cutoff
explicitly factorized out. Besides,
the operator in the matrix element is diagonal
in the basis, where $D$ and $D^*$ are diagonal,
\begin{equation}
\label{qnun}
\langle\,q|\,\nu,n\,\rangle\,=\,\frac 1{2\pi}
\bigl(q^2\bigr)^{-\frac 12+i\frac 12 \nu + i n\phi},
\end{equation}
$-\infty <\nu <\infty$, $n$ is integer,
$0<\phi< 2\pi$ is the polar angle in
the transverse $q$-space,
$$
\langle\nu,n\,|\,H(u,D,D^*)\,|\,\nu^\prime, n^\prime\rangle \,=\,
H(u,\nu,n)\,\delta(\nu-\nu^\prime)\,\delta_{n,n^\prime},
$$
$$
H(u,\nu,n)=H\bigl(u,D\to \frac 12 (-1+i\nu+n),
D^*\to \frac 12 (-1+i\nu-n)\bigr).
$$

From now on we suppose the impact factors
to be angular independent in the transverse space.
Then after inserting the complete set,
$$
\sum_n\int_{-\infty}^\infty d\nu
\langle\,q_2|\,\nu,n\,\rangle
\langle\,\nu,n|\,q_1\,\rangle\,=\,\delta^{(2)}(q_2-q_1),
$$
into the matrix element (\ref{P(u)f})
only the terms with $n=0$ survive in $P(u)$.
Suppose also $q_R$ is a typical momentum scale for the both
impact factors, $\Phi_{A,B}=\Phi_{A,B}(q^2/q_R^2)$ and introduce
the functions $C_{A,B}(n)$ according to the relation
\begin{equation}
\label{CAB}
\int \frac{d^2q}{q^2}\Phi_{A,B}(\frac{q^2}{q_R^2})\,
\bigl(q^2\bigr)^n\,=\,\bigl(q_R^2\bigr)^n\,C_{A,B}(n).
\end{equation}
With these notations one gets
\begin{equation}
\label{Pnu}
P(u)\,=\,e^{\,2L u\psi(1)}\left(\frac{q_R^2}{m_g^2}\right)^{-L\overline u}
\int d\nu\,
C_A\bigl(-\frac 12 L\overline u + \frac 12 i\nu\bigr)
C_B\bigl(-\frac 12 L\overline u - \frac 12 i\nu\bigr)\,
H(u,\nu,0).
\end{equation}
The function $H(u,\nu,0)$
embodies the scattering dynamics.
Taking the limit $\overline u \to 0$
we see that,
$$
P(1)\,=\,e^{\,2L \psi(1)}
\int\!\frac{d^{\,2}q}{q^2}\!\!
\int\!\frac{d^{\,2}l}{l^2}\,
\Phi_A(q)
\langle\,q|\,e^{K(1)L}|\,l\,\rangle\Phi_B(l)
$$
with
\begin{equation}
\label{K(1)}
K(1)\,=\,-\frac 12\bigl[\psi(1+D)+\psi(-D)
+ \psi(1+D^*)+\psi(-D^*)\bigr]
\end{equation}
and consequently $e^{\,2L \psi(1)}H(1,\nu,0)=e^{\omega(\nu)L}$,
where BFKL eigenvalue,
\begin{equation}
\label{omega}
\omega(\nu)\,=\,\alpha\bigl[2\psi(1)-\psi(\frac 12 + \frac 12 i\nu)
-\psi(\frac 12 - \frac 12 i\nu)\bigr],
\end{equation}
yields for $\nu=0$ the Pomeron mediated elastic cross section.
The main contribution for $L\gg 1$ comes to the integral
from the region near $\nu=0$,
$P(1)\sim C_A(0)C_B(0)\exp\bigl(\omega_P Y\bigr)$,
where $\omega_P=\omega(0)=4\alpha\ln 2$ is the LLA
Pomeron intercept.

The function $P(u)$ (\ref{Pnu}) turns out, however,
to be strongly suppressed when
$m_g\to 0$ at least for $u<1$.
On the other hand, the mean gluon number
$$
\overline N\,=\,
\frac 1{P(1)}\frac{\partial}{\partial u}P(u)\biggl|_{u=1}
\sim\, \ln\frac{q_R^2}{m_g^2},
$$
indicates the infinite growth of the emitted gluons
with decreasing infrared cut off $m_g$.
The reason behind this
is in the virtual gluons.
The function $P(u)$ is constructed in (\ref{Pnu}) to fix
the number of the real gluons whereas the virtual ones
associated with the gluon trajectory $\omega$ remain
unrestricted. As a consequence we are dealing with
the amplitudes of the given perturbation order for the real emission
allowing at the same time for the virtual corrections of an arbitrary order.
Different number of the real and virtual gluons, the latter number
being unbounded while the first one is finite, breaks down the infrared
cancelation.
A possible way
to improve it is to modify the operator $K(u)\to K(u,v)=u K_r+2v\,\omega$
by adding a new variable $v$ that "counts" the virtual gluons.
It simply results into redefinition $\overline u \to v-u$ in the all
above expressions, so that $u$ and $\overline u$ become independent.
Putting $m_g \to 0$ enforces $u=v$,
that leaves us with $\overline u \to 0$ limit,
$$
P(u)\,=\,C_A(0)C_B(0)\,e^{\,\omega_P Y u}.
$$
written here for the dominant, $\nu=0$, part.
Thus we arrive at
the Poisson distribution,
\begin{equation}
\label{PNf}
P_N\,=\,\frac 1{N!}\,(\omega_P Y)^N\,e^{-\omega_P Y}.
\end{equation}

This formula however is hardly directly applied
to the observable emission. The infrared cancelation of
real and virtual parts occurs in each perturbation order
that is why we actually get here the distribution over
the evolution steps rather than the emitted gluons.
Put somewhat differently, it is the distribution of the ladder
cells number in the Pomeron diagram for the exclusive cross section,
which is finite for $m_g\to 0$.
Dealing with inclusive processes one has to appreciate
that there is always a physical
infrared cut-off provided either by a sort of minimal
experimentally resolved momentum or some kinematical restrictions
imposed to select the proper observables.
Thus we have to modify our treatment to accomodate
the additional constraints.

\noindent
{\bf 3.} First, we modify the expression (\ref{P(u)f})
to get the distribution of the gluons produced with
a given total transverse momentum $Q$.
On this purpose the probability to emit $N$ gluons with transverse
momenta $k_i$ has to be multiplied by the factor
$$
\delta^{(2)}(Q - \sum_{i=1}^N k_i)\,=\,
\int\!\frac{d^2 z}{(2\pi)^2}\,e^{izQ}
\prod_{i=1}^N e^{-iz k_i}
$$
that results into modification of the real emission operator~(\ref{Kr}),
$$
\langle q|\,K_r\,|q^\prime\rangle \to
e^{-iz(q-q^\prime)}\langle q|\,K_r\,|q^\prime\rangle.
$$
It leads to the replacement
$\ln m_g^2x^2 \to \ln m_g^2(x-z)^2$ in the formula (\ref{log}).
Then the equalities
$$
\ln (x-z)^2\,=\,2\ln2\,+\,\ln\bigl(\frac{\partial}{\partial q}
+\frac i2 z^*\bigr)\,+\,\bigl[\ln\bigl(\frac{\partial}{\partial q}
+\frac i2 z^*\bigr)\bigr]^+,~~~~ z=z_1+iz_2,
$$
and the chain of relations
\begin{eqnarray}
\ln q\,+\,\ln\bigl(\frac{\partial}{\partial q}
+\frac i2 z^*\bigr)\,&=&\,
e^{-\frac 12 iz^*q}\,\bigl[\ln q+\ln\frac{\partial}{\partial q}
\,\bigr]e^{\frac 12 iz^*q} \nonumber \\
=\,e^{-\frac 12 iz^*q}\,\bigl[\frac 12
\psi\bigl(1+q\frac{\partial}{\partial q}\bigr)
+\frac 12\psi\bigl(-q\frac{\partial}{\partial q}\bigr)
\bigr]\,e^{\frac 12 iz^*q}\,
&=&\,\frac 12\bigl[\psi(1+D_z)+\psi(-D_z)\bigr],\nonumber
\end{eqnarray}
$$
D_z=q\,\bigl(\frac{\partial}{\partial q}+\frac i2z\bigr),
~~~~\bigl[\,\ln q\,,\,D_z\bigr]\,=\,-1
$$
together with its Hermitian conjugate
and the equality
$$
\langle\,q\,|H(u,D_z,D_z^*)|\,l\,\rangle
=\langle\,q\,|e^{-iqz}H(u,D,D^*)e^{iqz}|\,l\,\rangle
= e^{iz(-q+l)}\langle\,q\,|H(u,D,D^*)|\,l\,\rangle
$$
yield (\ref{P(u)f})
with the replacement
\begin{equation}
\label{dz}
\langle\,q\,|H(u,D,D^*)|\,l\,\rangle \to
\int\!\!\frac{d^{\,2} z}{(2\pi)^2}\,e^{iz(Q-q+l)}
\langle\,q\,|H(u,D,D^*)|\,l\,\rangle.
\end{equation}

The second modification
is the gap assumed
in the outgoing gluons momentum spectrum. We pick up the processes,
where all the gluons are produced
with the transverse momentum $q_i$ larger than a certain value $q_0$.
To fulfil the latter requirement one has to modify the real emission part
in (\ref{Kr}),
$\langle q|\,K_r\,|q^\prime\rangle \to
\theta\bigl((q-q^\prime)^2-q_0^2\bigr)\langle q|\,K_r\,|q^\prime\rangle$,
or, within logarithmic accuracy,
$\ln m_g^2x^2 \to \ln q_0^2x^2$ in the formula (\ref{log}).
The logarithmic accuracy implies the value $q_0$
to be small compared to the typical momentum
scale, $q_0^2\ll q_R^2$.
The power-like corrections beyond this region
make the relevant operators to be non diagonal
in the $|\nu,n\rangle$ basis, which invalidates
the method performance.

The third modification is that the secondaries are registered
not at all kinematically available rapidities $0<y<Y$
but in a more narrow range $0<y_1<y<y_2<Y$.
To implement this condition into generating function
one has to replace
$K \to K(u)$ (\ref{K(u)}) at the interval
$[y_1,y_2]$ to fix there the number of emitted particles,
whereas
the intervals $[0,y_1]$ and $[y_2,Y]$.
have to be left with unchanged BFKL operator $K=K(1)$ (\ref{K(1)}).
Thus instead of (\ref{P(u)}) one has
\begin{equation}
\label{PyYy}
P(u)\,=\,\int\!\frac{d^{\,2}q}{q^2}\int\!\frac{d^{\,2}l}{l^{\,2}}\,
\Phi_A(q)\langle\,q|\,e^{K(1)(Y-y_2)}e^{K(u)(y_2-y_1)}
e^{K(1)y_1}\,|l\,\rangle\,
\Phi_B(l).
\end{equation}

To proceed further, we first substitute the operator
in the middle with the above obtained expressions
(\ref{P(u)f}) and (\ref{HuDD}),
$$
\langle\,q|\,e^{K(1)(Y-y_2)}e^{K(u)(y_2-y_1)}
e^{K(1)\,y_1}|l\,\rangle
$$
$$
\,=\,
e^{\,L  u(2\psi(1)-\ln\frac{q_0^2}{m_g^2})}
\langle\,q|\,e^{K(1)(Y-y_2)}
\left(\frac{q^2}{m_g^2}\right)
^{\!\!-\frac 12L\overline u}
\!\!\! H(u,D,D^*)\,
\left(\frac{q^2}{m_g^2}\right)
^{\!\!-\frac 12L\overline u}
\!\!e^{K(1)y_1}\,|l\,\rangle.
$$
Here $L=\alpha(y_2-y_1)$.
The next step is purely algebraic.
With the help of identities
$$
D\,q^n\,=\,q^n\,(D+n),~~~~
D^*\,(q^*)^n\,=\,(q^*)^n\,(D^*+n),
$$
valid for any $n$,
we drag BFKL exponents with $K(1)$ operator to the right and
to the left through $q^2$ powers until they join the operator
$H(u,D,D^*)$,
$$
e^{K(1)(Y-y_2)}
\left(\frac{q^2}{m_g^2}\right)
^{\!\!-\frac 12L\overline u}
\!\! H(u,D,D^*)\,
\left(\frac{q^2}{m_g^2}\right)
^{\!\!-\frac 12L\overline u}
\!\! e^{K(1)\,y_1}
$$
$$
\,=\,
\left(\frac{q^2}{m_g^2}\right)
^{\!\!-\frac 12L\overline u}\!\!
e^{K_L(Y-y_2)}\,H(u,D,D^*)\,e^{K_R \,y_1}
\left(\frac{q^2}{m_g^2}\right)
^{\!\!-\frac 12L\overline u},
$$
$$
\begin{array}{c}
K_L\,=\,-\frac 12\bigl[\psi(1+D+\frac 12 L\overline u)
+\psi(-D-\frac 12 L\overline u)
+\psi(1+D^*+\frac 12 L\overline u)
+\psi(-D^*-\frac 12 L\overline u)
\bigr],\\
K_R\,=\,-\frac 12\bigl[\psi(1+D-\frac 12 L\overline u)
+\psi(-D+\frac 12 L\overline u)
+\psi(1+D^*-\frac 12 L\overline u)
+\psi(-D^*+\frac 12 L\overline u)
\bigr].
\end{array}
$$
The third step is to insert the complete set,
\begin{eqnarray}
P(u)\,&=&\,\sum\limits_{n}\int d\nu
\int\!\frac{d^{\,2}q}{q^2} \Phi_A(q)\langle\,q\,|
\left(\frac{q^2}{m_g^2}\right)
^{\!\!-\frac 12L\overline u}
\!\!|\nu,n\rangle  \\
&&\times
\langle\,\nu,n|\,e^{K_L(Y-y_2)}\,H(u,D,D^*)\
e^{K_R \,y_1}\,|\nu,n\,\rangle
\int\!\frac{d^{\,2}l}{l^{\,2}}\langle\,\nu,n|
\left(\frac{q^2}{m_g^2}\right)
^{\!\!-\frac 12L\overline u}
\!\!\!|\,l\rangle\,\Phi_B(l).\nonumber
\end{eqnarray}
Here we take into account that the operators $K_L,K_R,H$
are diagonal in $|\nu,n\rangle$ basis, besides
the angular independence
of the impact factors selects $n=0$ in the sum.
Recalling (\ref{qnun}), (\ref{CAB}) the momentum
integrals read
$$
\int\!\frac{d^{\,2}q}{q^2} \Phi_{A,B}(q)\langle\,q\,|
\left(\frac{q^2}{m_g^2}\right)
^{\!\!-\frac 12L\overline u}
\!\!\!|\nu,0\rangle\,=\,C_{A,B}(-\frac 12 L\overline u \pm \frac 12 i\nu)
(q_R^2)^{-\frac 12 \pm \frac 12 i\nu}
\left(\frac{q_R^2}{m_g^2}\right)
^{\!\!-\frac 12L\overline u}
$$
Thus we have
\begin{equation}
\label{omega0}
P(u)\,=\,\int d\nu \,e^{-L\ln\frac{q_R^2}{m_g^2}}\,
e^{L\,\omega_0\, u}\,T(u),
~~~~\omega_0=\ln\frac{q_R^2}{q_0^2}+2\psi(1),
\end{equation}
where
\begin{eqnarray}
\label{Tu}
T(u)\,&=&\,\int \! d\nu\,
C_{A}(-\frac 12 L\overline u + \frac 12 i\nu)\,
C_{B}(-\frac 12 L\overline u - \frac 12 i\nu)
 \\
&&\times\langle\,\nu,0|\,e^{K_L(Y-y_2)}\,H(u,D,D^*)\
e^{K_R y_1}\,|\nu,0\,\rangle. \nonumber
\end{eqnarray}
Notice that $u$-independent multiplies are irrelevant
in the generating function as they are absorbed in
its overall normalization. That is why the fictitious
gluon mass $m_g$ drops out of the final distribution.
The natural infrared cut off is provided by $q_0$ momentum.
In what follows we will assume the typical scale $q_R^2 \gg q_0^2$þ

If the value $\omega_0$
can be treated as a large parameter the function $P(u)$
results into Poisson distribution.
It clearly follows from (\ref{omega0}) by keeping there
the maximal $\omega_0$ power for each $N$, which amounts
to differentiating $e^{L\omega_0\,u}$ only
while the rest factors are taken just at $u=0$.
The mean gluon number then $\overline N = \omega_0L \gg 1$.
The first correction arises from the terms with
one power of $\omega_0$ less than in the main order,
\begin{eqnarray}
P_N\,&=&\,P_N^{(0)}\,+\,P_N^{(1)}\nonumber \\
&=&\,C\,
\frac 1{N!}\frac{\partial^N}{\partial u^N}P(u)\biggl|_{u=0}\,\approx\,
C\,\frac 1{N!}\bigl(\omega_0 L\bigr)^N\,T(0)\biggl[1\,+\,
N\frac 1{\omega_0 L}\,\frac{T^\prime(0)}{T(0)}\biggr]
\nonumber \\
&\approx&\,C\,\frac 1{N!}\bigl(\omega_0 L\bigr)^N\,T(0)\biggl[1\,+\,
\frac 1{\omega_0 L}\,\frac{T^\prime(0)}{T(0)}\biggr]^N \nonumber
\end{eqnarray}
where $C$ is the normalization constant. As is evident from
the last line the first order correction preserves
the form of the Poisson distribution,
$$
P_N\,=\,\frac 1{N!}\bigl(\omega_0^\prime L\bigr)^N\,
e^{-\omega_0^\prime L},
$$
affecting only the mean number of gluons,
$\overline N^\prime = \omega^\prime L=
\omega_0 L+\frac{T^\prime(0)}{T(0)}$.

We try to estimate
the first order contribution for $L\gg 1$,
supposing, in addition, both the rapidity intervals
to be equal (actually of the same order) and large,
$(Y-y_2)=y_1=y$, $\alpha y\gg 1$.

The leading for $L\gg 1$ terms could be expected
to arise from the function $H(u,\nu,0)$, therefore
the part of interest is
\begin{eqnarray}
\label{Tp}
T^\prime(0)&\approx&\int\! d\nu\,
C_{A}(-\frac 12 L + \frac 12 i\nu)\,
C_{B}(-\frac 12 L - \frac 12 i\nu)  \\
&&\times\exp\bigl\{\alpha y \bigr(K_L(0,\nu)+K_R(0,\nu)\bigl)\bigr\}
\frac{\partial}{\partial u}H(u,\nu,0)\bigr|_{u=0}.\nonumber
\end{eqnarray}
It is convenient to rewrite the sum
in the exponent as
$$
K_L(0,\nu)+K_R(0,\nu)\,=\,
-\psi\bigl(\frac 12+\frac 12 i\nu + \frac 12 L\bigr)
-\psi\bigl(\frac 12-\frac 12 i\nu + \frac 12 L\bigr)
-\frac{\pi\sin \pi L}{\cos\pi L +\cosh 2\pi\nu}.
$$
Aiming at the large $L$ asymptotics,
it is sufficient to take this expression only
for integer $L$, that removes at all the last term,
whereupon we have
$$
K_L(0,\nu)+K_R(0,\nu)\,=\,-2\psi\bigl(\frac 12 + \frac 12 L\bigr)
+\frac 14\,\psi^{\prime\prime}\bigl(\frac 12 + \frac 12 L\bigr)\,\nu^2\,
+\,O(\nu^4).
$$
Similarly presenting the function
\begin{eqnarray}
\label{dH}
\frac{\partial}{\partial u}H(u,\nu,0)\bigr|_{u=0}\,&=&\,
-2\ln\Gamma\bigl(\frac 12 + \frac 12 L -\frac 12 i\nu\bigr)
-2\ln\Gamma\bigl(\frac 12 + \frac 12 L +\frac 12 i\nu\bigr)
\\
&&-\ln\bigl(\cos \pi L + \cosh 2\pi\nu\bigr)+\ln 2\pi^2
\nonumber
\end{eqnarray}
makes it clear that only the first two terms
grow with $L$ while the last ones, being periodic
in $L$, do not and can be omitted.
Given now that $\psi^{\prime\prime}(x)\sim -1/x^2$ for $x\gg 1$
the typical $\nu$ values in the $T(0)$ integral (\ref{Tu}),
$\nu^2 \sim L^2/(\alpha y)$, are small compared to $L^2$ when $\alpha y\gg 1$.
It allows to neglect $\nu$ in the function
$H^\prime(0,\nu,0)$
as well as in the impact factor functions $C_{A,B}$ in (\ref{Tp}),
that finally yields
\begin{equation}
\frac{T^\prime(0)}{T(0)}\,=\,-L\ln\frac{L}{2}+L\,+\,O(1).
\end{equation}

It is worth to point out that in terms
of the original basic decomposition
(\ref{P(u)f}) the main contribution to the multiplicity
is due to the power multipliers
whereas the function $H(u,D,D^*)$ specifies the corrections.

The generating function for the distribution of the gluons
carrying in aggregate the fixed total transverse momentum $Q$
according to (\ref{P(u)f}) and (\ref{dz}) is
\begin{eqnarray}
P(Q,u)&=&\int\!\!\frac{d^{\,2} z}{(2\pi)^2}\,
e^{\,L  u(2\psi(1)-\ln\frac{q_0^2}{m_g^2})}
e^{izQ}\,P(z,u),\nonumber \\
\label{Pz}
P(z,u)&=&\int\!\frac{d^{\,2}q}{q^2}\!\!\int\!\!\frac{d^{\,2}l}{l^{\,2}}\,
\Phi_A(q)\left(\frac{q^2}{m_g^2}\right)
^{\!\!-\frac 12L\overline u}
\!\!\!\langle\,q|e^{-iqz}H(u,D,D^*)e^{iqz}\,|l\,\rangle
\!\left(\frac{l^2}{m_g^2}\right)
^{\!\!-\frac 12L\overline u}
\!\!\!\Phi_B(l).\nonumber
\end{eqnarray}
The momentum $Q$ is the total momentum of the gluons
that are supposed to be registered, that is only those
emitted with the transverse momenta larger than $q_0$.

There is an important difference from the previous case,
where the main order source is in the terms with
no derivatives of the function $H(u,D,D^*)$.
Here the dominant terms including $H(0,D,D^*)=1$ result
into $z$-independent contribution that in turn
produces $P(Q,u)\sim \delta^{(2)}(Q)$.

To find the leading behavior at small $q_0$
and the total momentum $Q\not =0$ we apply the formal
trick:
$$
H(u,D,D^*)\,=\,H(u,\frac{\partial}{\partial t},
\frac{\partial}{\partial t^*})e^{t\,D}e^{t^*D^*}|_{t=t^*=0},
$$
and employ the operator identities
$$
\begin{array}{lr}
  e^{t\,D}e^{t^*D^*}\,e^{iqz}\,=\,e^{i\overline q z}
e^{t\,D}e^{t^*D^*}, &
\overline q_1=\frac 12(t+t^*)q_1 -\frac 1{2i}(t-t^*)q_2,
 \\
  & \overline q_2=\frac 1{2i}(t-t^*)q_1 +\frac 1{2}(t+t^*)q_2,
\end{array}
$$
$$
e^{t\,D}e^{t^*D^*}\!\int d^2 l\,|\,l\rangle \frac 1{l^2}
\Phi_B\bigl(\frac{l^2}{q_R^2}\bigr)\,=\,
\int d^2l\,|\,l\rangle
\frac{e^{-t-t^*}}{l^2}
\Phi_B\bigl(e^{t+t^*}\frac{l^2}{q_R^2}\bigr).
$$
Then we get
\begin{eqnarray}
P(z,u)\,&=&\,H(u,\frac{\partial}{\partial t},
\frac{\partial}{\partial t^*})\biggl|_{t=t^*=0}~~
e^{-(t+t^*)(\frac 12 L\overline u+1)}
\nonumber \\
&&\times\,\int \frac{d^2q}{q^4}
e^{-i(q - \overline q)z}
\left(\frac{q^2}{m_g^2}\right)
^{\!\!-L\overline u}
\!\Phi_A\bigl(\frac{q^2}{q_R^2}\bigr)\,
\Phi_B\bigl(e^{t+t^*}\frac{q^2}{q_R^2}\bigr)
\nonumber
\end{eqnarray}
or doing $z$ integrals, that return
$(2\pi)^2\delta^{(2)}(Q-q+\overline q)$,
\begin{eqnarray}
P(Q,u)&=&e^{-L\ln\frac{Q^2}{m_g^2}}\,
e^{L(2\psi(1)+\ln\frac{Q^2}{q_0^2})u}
\,H(u,\frac{\partial}{\partial t},
\frac{\partial}{\partial t^*})\biggl|_{t=t^*=0}
~~e^{-(t+t^*)(\frac 12 L\overline u+1)}
~n_{t,t^*}^{L\overline u -1}
\nonumber \\
&&\times\,\frac 1{Q^4}
\Phi_A\bigl(\frac 1{n_{t,t^*}}\,\frac{Q^2}{q_R^2}\bigr)\,
\Phi_B\bigl(\frac{e^{t+t^*}}{n_{t,t^*}}\,
\frac{Q^2}{q_R^2}\bigr),
~~~~ n_{t,t^*}\equiv (1-e^t)(1-e^{t^*}).\nonumber
\end{eqnarray}
In principle this form could offer one more tool
to handle the generating function, but here it is only needed
to factorize out the leading asymptotics
assuming $\omega_0(Q)=2\psi(1)+\ln\frac{Q^2}{q_0^2} \gg 1$.
The non vanishing result is obtained after
the function $H$ is once differentiated,
$$
P_N^{(1)}(Q)\,=\,e^{-L(\ln\frac{q_R^2}{m_g^2}+\ln\frac{Q^2}{q_R^2})}
\,C_1\bigl(\frac{Q^2}{q_R^2}\bigr)\,\frac 1{(N-1)!}\,
\omega_0^{N-1}(Q).
$$
Although this is a first term
it looks like a correction to the Poisson distribution,
the computation of $C_1$ function being
similar to what has been done above for $T^\prime(0)$.
The analog of the leading term
comes from the delta-function piece,
$$
P_N^{(0)}(Q)\,=\,e^{-L(\ln\frac{q_R^2}{m_g^2}}\,C_0\,
\frac 1{N!}\,\bigl(\omega_0 L\bigr)^N\,\delta^{(2)}(Q),~~
C_0\,=\,\frac 1{q_R^2}\int d^2x\,\Phi_A(x)\Phi_B(x)
\bigl(x^2\bigr)^{-L-2}.
$$
If the total momentum essential range is $q_0^2\ll Q^2 \ll q_R^2$
(due to impact factors) the value
$C_1\bigl(\frac{Q^2}{q_R^2}\bigr)\approx C_1(0)$
could be found from the relation
$$
\int d^2Q\,\bigl[P_N^{(0)}(Q) + P_N^{(1)}(Q)\bigr]\,=\,
e^{-L\ln\frac{q_R^2}{m_g^2}}\biggl[
\frac 1{N!}\bigl(\omega_0 L\bigr)^N T(0)\,+\,
\frac 1{(N-1)!}\bigl(\omega_0 L\bigr)^{N-1} T^\prime(0)
\biggr]
$$
for known $T(0)$, $T^\prime(0)$.

\noindent
{\bf 4.} Concluding,
the distribution of the gluons arising from the cut
of BFKL Pomeron in LLA has been found for the two separate cases.
The first one is entirely unobservable although it literally
refers to the gluons "inside" the Pomeron, that is to
the distribution of the ladder diagrams in
the Pomeron Green function. It is exactly
of the Poisson type,
the "mean number" of the gluons, or the ladder "length",
being proportional to the Pomeron intercept.

The second case concerns the real
multiple emission in the scattering. The gluon distribution
is generally more complex
but getting closer to the Poisson one, when the gap parameter $q_0$,
that plays the role of the infrared cut off, decreases.
However in this case the mean gluon number turns out
not to be identical to the intercept.

The first order corrections to the Poisson
distribution obtained for $L=\alpha Y \gg 1$
affects only its mean number parameter.
The distribution of the gluons with a fixed total transverse
momentum looks like the first order correction with
the main term vanishing for the non zero momentum value.

The next order corrections as well as the effects from $L\sim 1$
presumably cause the deviation from
the Poisson distribution but its form would be more
dependent on the impact factors. Here we are mainly
interested in the effects of BFKL dynamics and have used
roughly estimated impact factors.

This situation looks different compared to $e^+e^-$ case,
where more complicated parton distributions are
obtained~\cite{Dokshitzer:1993dc}.
Assuming soft branching hypotheses~\cite{Dokshitzer:1982ia}
telling that main qualitative features of the secondary
hadrons distribution are similar to those of
the partons one plausibly expects the secondary hadron
to be distributed similarly to the gluons
according to Poisson form.
It could explain the observed KNO scaling
effects~\cite{KNO, Kaidalov:1982xe}.

The authors are grateful to M.G.~Ryskin for helpful discussion.


\begin{thebibliography}{**}

\bibitem{KNO}
Z. Koba, H.B. Nielsen and P. Olesen,
Nucl. Phys. {\bf B40} (1972) 372.

\bibitem{alner}
G.J. Alner  et al.  (UA5 Collaboration),
Phys.\ Lett.\  {\bf B138 } (1984) 304.

\bibitem{ansorge}
R.E. Ansorge et al.  (UA5 Collaboration),
Z.\ Phys.\  {\bf C43}, (1989)  357

\bibitem{CMV}
M. Ciafaloni, G. Marchesini, G. Veneziano,
Nucl.Phys. {\bf B98} (1975) 472.

\bibitem{Kaidalov:1982xe}
A.~B.~Kaidalov and K.~A.~Ter-Martirosian,
Phys.\ Lett.\  {\bf 117B} (1982) 247.

\bibitem{Abramovsky:1973fm}
V.~A.~Abramovsky, V.~N.~Gribov and O.~V.~Kancheli,
Yad.\ Fiz.\  {\bf 18} (1973) 595
[Sov.\ J.\ Nucl.\ Phys.\  {\bf 18} (1974) 308].

\bibitem{Kaidalov:1983vn}
A.~B.~Kaidalov and K.~A.~Ter-Martirosian,
Sov.\ J.\ Nucl.\ Phys.\  {\bf 39} (1984) 979
[Yad.\ Fiz.\  {\bf 39} (1984) 1545].

\bibitem{Kaidalov:1983ew}
A.~B.~Kaidalov and K.~A.~Ter-Martirosyan,
Sov.\ J.\ Nucl.\ Phys.\  {\bf 40} (1984) 135
[Yad.\ Fiz.\  {\bf 40} (1984) 211].

\bibitem{Hegyi:2000sp}
S.~Hegyi,
Nucl.\ Phys.\ Proc.\ Suppl.\  {\bf 92} (2001) 122
[hep-ph/0011301].

\bibitem{Dokshitzer:1993dc}
Y.~L.~Dokshitzer,
Phys.\ Lett.\ B {\bf 305} (1993) 295.

\bibitem{Dokshitzer:1991wu}
Y.~L.~Dokshitzer, V.~A.~Khoze, A.~H.~Mueller and S.~I.~Troian,
{\em Basics of perturbative QCD},
Gif-sur-Yvette, France: Ed. Frontieres (1991) 274 p.

\bibitem{FKL}
V. S. Fadin, E. A. Kuraev and L. N. Lipatov,
Multi-Regge form of many-particle amplitudes
Phys. Lett. {\bf B 60} (1975) 50.

\bibitem{KLF1}
E. A. Kuraev, L. N. Lipatov and V. S. Fadin,
Zh. Eksp. Teor. Fiz. {\bf 71} (1976) 840 [Sov.
Phys. JETP {\bf 44} (1976) 443].

\bibitem{KLF2}
E. A. Kuraev, L. N. Lipatov and V. S. Fadin,
Zh. Eksp. Teor. Fiz. {\bf 72} (1977) 377 [Sov.
Phys. JETP {\bf 45} (1977) 199].

\bibitem{BL}
Balitsky, I.I. and Lipatov, L. N.,
Sov. J. Nucl. Phys. {\bf 28} (1978) 822
[Yad. Fiz. {\bf 28} (1978) 1597].

\bibitem{FF}
Fadin, V. S. and Fiore, R. Phys. Lett. {\bf B440}
(1998) 359.

\bibitem{Fadin:2006bj}
V.~S.~Fadin, R.~Fiore, M.~G.~Kozlov and A.~V.~Reznichenko,
Phys.\ Lett.\ B {\bf 639}, 74 (2006)
[hep-ph/0602006].

\bibitem{deVega:2003ji}
H.~J.~de Vega and L.~N.~Lipatov,
Phys.\ Lett.\ B {\bf 578} (2004) 335
[hep-ph/0310124].

\bibitem{IFL}
B. L. Ioffe, V. S. Fadin and L. N. Lipatov,
Quantum Chromodynamics
Perturbative and Nonperturbative Aspects,
Cambridge University Press, 2010, ISBN:
9781107424753

\bibitem{Fadin:2015zea}
V.~S.~Fadin, M.~G.~Kozlov and A.~V.~Reznichenko,
Phys.\ Rev.\ D {\bf 92} (2015) no.8,  085044
[arXiv:1507.00823 [hep-th]].

\bibitem{Dokshitzer:1982ia}
Y.~L.~Dokshitzer, V.~S.~Fadin and V.~A.~Khoze,
Z.\ Phys.\ C {\bf 18} (1983) 37.

\end{thebibliography}
\end{document}